\documentclass[twocolumn,showpacs,preprintnumbers,amsmath,amssymb]{revtex4}

\usepackage{graphicx}
\usepackage{dcolumn}
\usepackage{bm}

\begin{document}
\title{Domain wall dynamics driven by adiabatic spin transfer torques \\
}

\author{Z. Li and S. Zhang}
\affiliation{ Department of Physics and Astronomy, University of
Missouri-Columbia, Columbia, MO 65211 }

\date{\today}

\begin{abstract}

In a first approximation, known as the adiabatic process, the
direction of the spin polarization of currents is parallel to the
local magnetization vector in a domain wall. Thus the spatial
variation of the direction of the spin current inside the domain
wall results in an adiabatical spin transfer torque on the
magnetization. We show that domain wall motion driven by this spin
torque has many unique features that do not exist in the
conventional wall motion driven by a magnetic field. By
analytically and numerically solving the Landau-Lifshitz-Gilbert
equation along with the adiabatic spin torque in magnetic
nanowires, we find the domain wall has its maximum velocity at the
initial application of the current but the velocity decreases to
zero as the domain wall begins to deform during its motion. We
have computed domain wall displacement and domain wall deformation
of nanowires, and concluded that the spin torque based on the
adiabatic propagation of the spin current in the domain wall is
unable to maintain wall movement. We also introduce a novel
concept of domain wall inductance to characterize the capacity of
the spin-torque induced magnetic energy stored in a domain wall.
In the presence of domain wall pinning centers, we construct a
phase diagram for the domain wall depinning by the combined action
of the magnetic field and the spin current.
\end{abstract}


\maketitle
\section{introduction}
The subject of current-induced magnetization reversal (CIMR) in
magnetic multilayers has received considerably interest recently
\cite{Buhrman,Fert1,Kent,Bass1,Tsoi,Myers}. From fundamental point
of view, this topic introduces a new interaction between
non-equilibrium conduction electrons and local moments, and the
physics of this new interaction has not been well explored. From
the application point of view, the current-induced magnetization
reversal opens a novel way to control and manipulate the
magnetization dynamics that is one of the central issues in modern
magnetic technologies. In conventional magnetic devices, the
magnetization direction is controlled by external magnetic fields
generated by a current or by a permanent magnet. These magnetic
fields are usually spread into a relatively large spatial extent.
The CIMR can be confined to an exact spatial region where the
current flows; this property is very attractive for magnetic
nano-devices, e.g., magnetic random access memory.

The physics of the current-induced magnetization reversal involves
interplay between non-equilibrium conduction electrons and local
magnetization. Namely, the spin angular momentum carried by spin
polarized electrons is transferred to the local moment through the
exchange interaction. This is somewhat analogue to the well-known
phenomenon of electromigration \cite{Sorb} where (impurity) atoms
move along the direction opposite to the direction of the current.
The origin of the electromigration is mainly from the transfer of
{\em linear momentum} of non-equilibrium conduction electrons to
atoms, resulting a ``wind force'' which drives migration of the
atom. For magnetic materials where the electric current is spin
polarized, it is possible that the spin-polarization of
non-equilibrium conduction electrons changes its orientation along
the direction of the current, and thus the localized magnetic
moment receives a spin torque if the conduction electron
continuously transfers their spin angular momentum to the local
moment in the steady state of the current flow. This
non-equilibrium conduction electron induced spin torque can alter
magnetization structures, drive magnetization dynamics and even
create the domain wall motion.
\begin{figure}
\centering
\includegraphics[width=6.5cm]{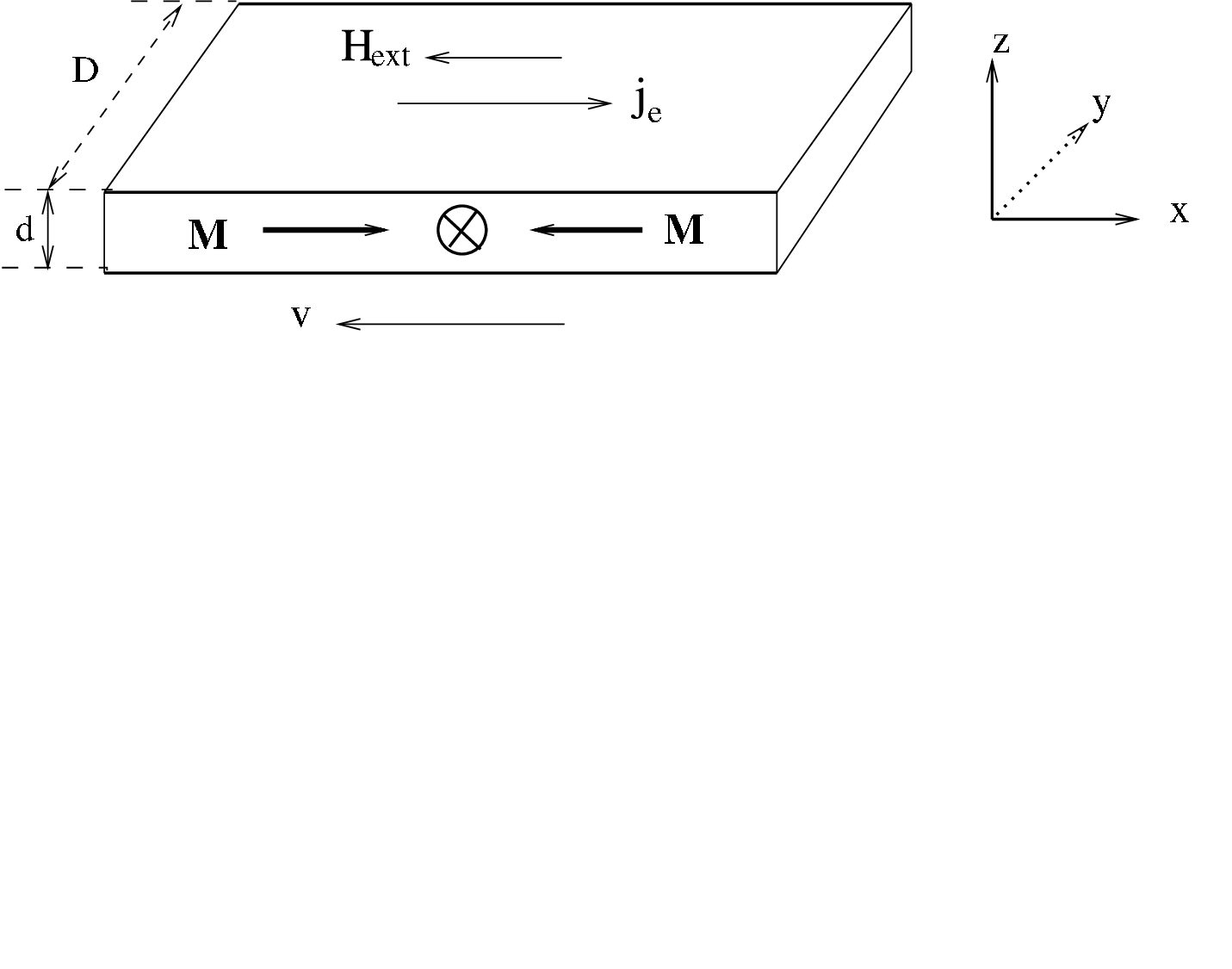}
\caption{Cartesian and polar coordinate systems}
\end{figure}
Up until now, the spin transfer model has been mainly on a
trilayer structure where two ferromagnetic layers are separated by
a spacer layer. Typically, one chooses the spacer layer thick
enough to reduce the magnetic coupling between the two
ferromagnetic layers and thin enough to minimize the spin-flip
scattering in the spacer so that the spin transfer torque is
maximized. Several theoretical models
\cite{Slon,Waintal,Stiles2,Brataas,Levy} have been put forward to
formulate the spin torque in terms of geometric and materials
parameters. While it is still debatable on the correct microscopic
picture of the spin torque, all theories contain a spin torque in
the form of
\begin{equation}
\mbox{\boldmath $\tau$}_a = - \frac{a_J}{M_s} {\bf M} \times ({\bf
M} \times \hat{\bf M}_p)
\end{equation}
where ${\bf M}$ is the magnetization of the free layer, $\hat{\bf
M}_p$ is the {\em unit} vector along the direction of the
magnetization of the pinned layer, $M_s$ is the saturation
magnetization, and $a_J$ is a model-dependent parameter and is
proportional to the current density. In general, the magnitude $a_J$
of the spin transfer torque is also a function of the angle between the
pinned and free layer magnetization vectors.

Several theoretical attempts \cite{Sun2,Bazaliy2,Miltat3,Li2} were
made in understanding the unique features displayed in the spin
torque, Eq.~(1). Here we just list three of them below: 1) the
spin torque may stabilize steady precessional states, i.e., the
magnetization does not converge to a meta-stable static state even
at zero temperature; 2) the spin torque may create a new stable
configuration of the magnetization, e.g., an out-of-plane
magnetization direction can be a new solution of meta-stable
states; 3) the spin torque modifies the effective energy barrier
or temperature in a significant and unique way.

All of the above novel properties are derived based on the
specific form of the spin torque, Eq.~(1). Equation (1) has
assumed that both free and pinned layers have uniform
magnetization within the layers, at least along the direction of
the current. In a typical ferromagnet, the magnetization is rarely
uniform and the dynamical process of magnetization is not coherent
rotation in general. It would be interesting to find what the spin
torque does to magnetization dynamics of non-single domain
ferromagnets. Berger \cite{Berger} introduced the ``domain drag
force'' by considering the spin torque in a single layer system
where the magnetization is not uniform along the current. He
argued, based on his intuitive physics picture, that the current
can drag the domain wall along the path of the current flow.
Bazaliy {\em et al.} proposed a spin torque in a ferromagnet
within the ballistic transport model for half-metallic materials
\cite{Bazaliy}. Most recently, we have generalized the spin torque
for any diffusive transport ferromagnet \cite{Zhang},
\begin{equation}
\mbox{\boldmath $\tau$}_{st} = - b_{J} \hat{\bf M} \times [
\hat{\bf M} \times (\hat{\bf j}_e \cdot \mbox{\boldmath $\nabla$}
) {\bf M} ]
\end{equation}
where $\hat{\bf M}= {\bf M}/M_s$, ${\hat{\bf j}}_e={\bf j}_e/j_e$
and $b_J= Pj_e \mu_B/eM_s$, $P$ is the spin polarization of the
current, $\mu_B$ is Bohr magnetron, and $ j_e$ is the electric
current density. The above torque has identical form as that of
Bazaliy {\em et al} \cite{Bazaliy}. However, we derived the above
formulation on a general property of ferromagnetic materials: in a
ferromagnet, the spin polarization of the current is always along
the direction of the local magnetization vector, i.e., the
transverse component of the spin current can be neglected
\cite{Stiles2,Brataas,Zhang}. Thus one may define a spin current
tensor ${\cal J} = (\mu_B/eM_s) P {\bf j}_e \otimes {\bf M}({\bf
r})$. The spin transfer torque is defined as the rate change of
the angular momentum of conduction electrons that are transferred
to the local magnetization, $\mbox{\boldmath $\tau$}_{st} = d{\bf
m}/dt = \partial{\bf m}/\partial t + \nabla \cdot {\cal J}$. In
the steady state of the current flow, $\partial{\bf m}/\partial t
= 0$ and thus we find the above spin torque can be written as
Eq.~(2) if we use the fact that the magnitude of the local
magnetization vector is a constant.

While the magnitude of the spin torque in spin valves, Eq.~(1),
depends on many unknown parameters such as interface
spin-dependent resistance, the magnitude of the adiabatic spin
transfer torque, $b_J$ in Eq.~(2), can be accurately estimated for
many ferromagnetic materials. $b_J$, which has a dimension of
velocity, is determined by two materials parameters: $M_s$ and
$P$; these parameters have already been determined experimentally.
In table I, we list the values of $b_J$ for a selected set of
materials for a spin current density $j_{e}=10^{7}A/cm^{2}$. The
half-metals, $CrO_2$ and $Fe_2O_3$, has a maximum spin
polarization (P=1) and has low saturation magnetization, and thus
the spin torque is larger than that of transition metals.
\begin{center}{{Table I The values of $b_{J}$ for some materials
for $j_{e}=10^{7}A/cm^{2}$}}
{
\begin{small}
\tabcolsep0.2in
\begin{tabular}{|c|c|c|c|}
\hline
Nanowire       &   $M_{s} (A/m)$    &  P   &   $b_{J} (m/s)$\\
\hline
Fe   &  $17.18\times10^{5}$  &0.5&1.35\\
\hline
Co & $14.46\time10^{5}$ & 0.35 &1.41\\
\hline
Ni & $4.9\times10^{5}$ & 0.23&2.7\\
\hline
Permalloy & $8\times10^{5}$ & 0.7&5.1\\
\hline
$\gamma-Fe_{2}O_{3}$ & $4.14\times10^{5}$ & 1.0 &14.0\\
\hline
$CrO_{2}$ & $3.98\times10^{5}$ & 1.0& 14.6\\
\hline
\end{tabular}
\end{small}
}
\end{center}

The paper is organized as follows. In Sec.~II,
we establish the equation of motion of a domain
wall by using the generalized Landau-Lifshitz-Gilbert equation along
with the adiabatic torque, Eq.~(2). We write the equation for nanowires. In
Sec~III, we propose a trial function to analytically obtain
solutions for domain wall dynamics. In particular, we address the
spin torque induced domain propagation and distortion. Then a
detailed comparison between numerical results and analytic
solutions is made. Finally, we summarize our major findings in
Sec. V.

\section{Model system}

We assume the current flows in $x$-direction, along the long
length of a wire, i.e., $\hat{\bf j}_e = {\bf e}_x$. By placing
this into Eq.~(2) and by treating Eq.~(2) as an additional torque
on the standard LLG equation, we write the generalized LLG
equation in the presence of the spin torque as,
\begin{equation}
\label{LLG} \frac{\partial {\bf M}}{\partial t}=-\gamma {\bf
M}\times {\bf H}_{eff}+\alpha{\hat{\bf M}} \times \frac{\partial
{\bf M}}{\partial t}-b_{J}{\hat{\bf M}}\times\left({\hat{\bf
M}}\times \frac{\partial{\bf M}}{\partial x}\right)
\end{equation}
where $\gamma$ is the gyromagnetic ratio, and ${\bf H}_{eff}$ is
the effective magnetic field including the external field, the
anisotropy field, magnetostatic field, and the exchange field,
and $\alpha$ is the Gilbert damping parameter.

We now specify the geometry of the wire and the effective field
entering Eq.~(3). As shown in Fig.~1, the $x$-axis is taken as the
easy axis as well as the direction of the external field
$H_{ext}$. The width and the thickness of the wire are $D$ and
$d$, respectively. A current is along the $x$ direction.
We explicitly write the effective field,
\begin{equation}
{\bf H}_{eff} = \frac{H_K M_x}{M_s} {\bf e}_x + \frac{2A}{M_s^2}
\nabla^2 {\bf M} - 4 \pi M_z {\bf e}_z + H_{ext} {\bf e}_x
\end{equation}
where $H_K$ is the anisotropy field, $A$ is the exchange constant,
and $4 \pi M_z$ is the de-magnetization field.

To analytically gain insight on the domain wall dynamics, we first
consider that the magnetization ${\bf M}$
varies only in the direction of $x$-axis. By placing Eq.~(4) into (3), the LLG
equation can be conveniently written in the polar coordinate,
\begin{eqnarray}
\frac{\partial \theta}{\partial t}+\alpha \sin\theta
\frac{\partial \varphi}{\partial t}=\gamma
\frac{2A}{M_s}\left(2\cos\theta \frac{\partial \varphi}{\partial
x}\frac{\partial \theta}{\partial x}+\sin\theta
\frac{\partial^{2}\varphi}{{\partial x}^{2}}\right)\nonumber\\
-4 \gamma \pi M_s \sin\theta\sin\varphi\cos\varphi +b_{J}
\frac{\partial
\theta}{\partial x}\\
\alpha \frac{\partial \theta}{\partial t}-\sin\theta
\frac{\partial \varphi}{\partial t}=\gamma\frac{2A}{M_s}
\left(\frac{\partial^{2}\theta}{{\partial
x}^{2}}-\sin\theta\cos\theta \left(\frac{\partial
\varphi}{\partial
x}\right)^{2}\right) \nonumber \\
-\gamma H_K \sin\theta \cos\theta -4\gamma \pi M_s
\sin\theta\cos\theta \sin^{2}\varphi\nonumber\\
-\gamma H_{ext}\sin\theta-b_{J} \sin\theta \frac{\partial
\varphi}{\partial x}
\end{eqnarray}
where $\theta$ represents the angle between the magnetization
vector and $x$-axis, and  $\varphi$ is the out-of-plane angle of
the magnetization vector projected in the $yz$ plane, i.e., $M_z =M_s
\sin\theta \sin\varphi $.

Our goal is to solve the magnetization vector as a function of
position $x$ and time $t$ from the above equations of motion. In
the absence of a current and/or an external magnetic field, we
consider that the domain wall, separated by two head-to-head
domains along the wire direction, is a N\'{e}el wall whose
magnetization stays in the plane of the wire, i.e., $M_z =0$,
$M_x=M_s \cos\theta$, $M_y = M_s \sin \theta\cos\varphi $, where
\begin{eqnarray}
\varphi=0; \hspace{0.3in} \theta = 2 {\rm tan}^{-1} \exp(x/W_0)
\end{eqnarray}
and $W_0 =\sqrt{2A/H_{K}M_s}$ is the domain wall width. At $t=0$, an
electric current/an external field is applied. We should determine
the motion of the wall at $t>0$ from Eqs.~(5) and (6) by utilizing the
initial condition, Eq.~(7).

\section{Analytical results}

The non-linear partial-differential equations, Eqs.~(5) and (6),
are difficult to solve. Here we follow Walker's analysis of domain
wall motion by introducing a trial function \cite{Walker}.
\begin{eqnarray}
\varphi=\varphi({\it t}); \hspace{0.3in}
\ln\tan\frac{\theta}{2}=c({\it t})\left(x-\int_{0}^{\it t}{\it
v}(\tau) d \tau\right) \, .
\end{eqnarray}
The first equation assumes that the projection of the
magnetization vector in the domain wall on the $yz$ plane is
independent of the position. One should note that the spatial
independence of $\varphi({\it t})$ does not mean a uniform
out-of-plane component because $M_z = \sin\theta \sin \varphi$ and
$\theta$ varies spatially in the domain wall from $\theta=0$ to
$\pi$. The second equation in Eq.~(8) postulates that the domain
wall shape remains a standard N\'{e}el-wall form except that the
width of the wall $W(t) \equiv c({\it t})^{-1}$ varies with time
and the wall moves with an instantaneous velocity ${\it v}({\it
t})$. As we will show later that this form of the solution is
indeed a correct solution as long as the spin torque $b_J$ and the
external field $H_{ext}$ are small.

By placing this trial function into Eqs.~(5) and (6), and by utilizing
the following identities,
\begin{eqnarray}
\frac{\partial \varphi}{\partial x} = \frac{\partial^{2} \varphi}
{{\partial x}^{2}}=0 \nonumber \\
\frac{\partial \theta}{\partial x}=c({\it t})\sin\theta;
\hspace{0.3in} \frac{\partial^{2} \theta}{\partial x^{2}}=c({\it
t})^{2}\sin\theta\cos\theta   \nonumber \\
\frac{\partial \theta}{\partial {\it t}}=c_{1}(x,
{\it t})\sin\theta \nonumber
\end{eqnarray}
and defining a function $c_{1}(x,{\it t})$,
\begin{equation}
c_{1}(x,{\it t}) \equiv \frac{d c({\it t})}{d {\it
t}}\left(x-\int_{0}^{\it t}{\it v}(\tau) {\it d} \tau
\right)-c({\it t}){\it v}({\it t})
\end{equation}
we find
\begin{eqnarray}
c_{1}(x,{\it t})+\alpha\frac{d\varphi}{d{\it t}}=-4\pi\gamma
M_s\sin\varphi\cos\varphi+b_J c({\it t})\\
-\alpha c_{1}(x,{\it t})+\frac{d \varphi}{d{\it t}}-\gamma
H_{ext}=\nonumber\\
\gamma\left(H_K-\frac{2A}{M_s}c({\it
t})^{2}+4\pi M_s\sin^{2}\varphi\right)\cos\theta
\end{eqnarray}
Strictly speaking, Eqs.~(10) and (11) can not be valid because
$c_1 (x,t)$ defined in Eq.~(9) is linearly proportional to $x$ but
there are no other terms in Eq.~(10) and (11) containing spatial
variable $x$. Thus, the trial function of the form of Eq.~(8) is
only possible when one discards the spatial dependence of $c_1$.
Equivalently, the first term in the right side of Eq.~(9) must be
much smaller compared to the second term so that $c_1 \approx -
c(t) v(t)$. While one would not determine the smallness of this
first term {\em a prior}, we will later confirm that $c(t)$ only
shrinks by a few percent during the domain wall motion and it
varies slowly with time so that $dc(t)/dt$ is indeed small. Now
the left-hand side of Eq.~(11) becomes $x$-independent and thus
the prefactor in front of $\cos \theta $ at the right-hand side of
Eq.~(11) must be identically zero,
\begin{equation}
H_K-\frac{2A}{M_s}c({\it t})^{2}+4\pi
M_s\sin^{2}\varphi=0
\end{equation}
and
\begin{equation}
\alpha c(t) v(t) +\frac{d \varphi}{d{\it t}}=\gamma H_{ext} \, .
\end{equation}
By placing Eq.~(13) into (10), we have
\begin{equation}
(1+\alpha^{2})\frac{d\varphi}{d{\it t}}=\gamma H_{ext}+\alpha
b_{J} c({\it t})-4\pi\alpha\gamma M_s\sin\varphi\cos\varphi \, .
\end{equation}
Equations (12) and (14) are the ordinary first-order differential
equations that determine the domain wall width $c^{-1}(t)$ and the
rotation of the domain wall plane $\varphi (t)$ subject to the
initial values $\varphi (t=0)=0$ and $c^{-1}(t=0) = 1/W_0 $. Once
these two equations are solved, we can obtain the velocity of the
domain wall from Eq.~(13)
\begin{equation}
{\it v}({\it t})=\frac{\gamma(\alpha H_{ext}+4\pi
M_s\sin\varphi\cos\varphi)}{(1+\alpha^2)c({\it
t})}-\frac{b_{J}}{1+\alpha^2} \, .
\end{equation}
The numerical solutions of Eq.~(12) and (14) are represented in
Fig.~2 where the domain wall velocity, displacement, and
distortion as a function of time are shown. In the following
subsections, we discuss these quantities.

\subsection{Domain wall velocity}

At t=0, the wall is a standard in-plane N\'{e}el wall, i.e.,
$\varphi (t=0) =0$. From Eq.~(15), we have readily seen that the
velocity is $-b_J/(1+\alpha^2)$ at the initial application of the
current if no external field is applied. On the other hand, for
${\it t}\rightarrow \infty$, $d\varphi/d{\it t}=0$. We immediately
find, from Eq.~(13),
\begin{equation}
{\it v}_{s} \equiv v( \infty ) = \frac{\gamma H_{ext}}{c(\infty)
\alpha};
\end{equation}
this result is same as that of Walker \cite{Walker} in the
absence of the spin torque. We conclude that the terminal velocity is
independent of the spin torque at all! This implies that the spin current alone,
i.e., no applied field, can not move the domain wall to a large distance.
In Fig.2(a), the velocity of the domain wall during the application
of the current is shown. It is noted that
the domain wall motion stops at a fraction of a nanosecond.

\begin{figure}
\centering
\includegraphics[width=7.5cm]{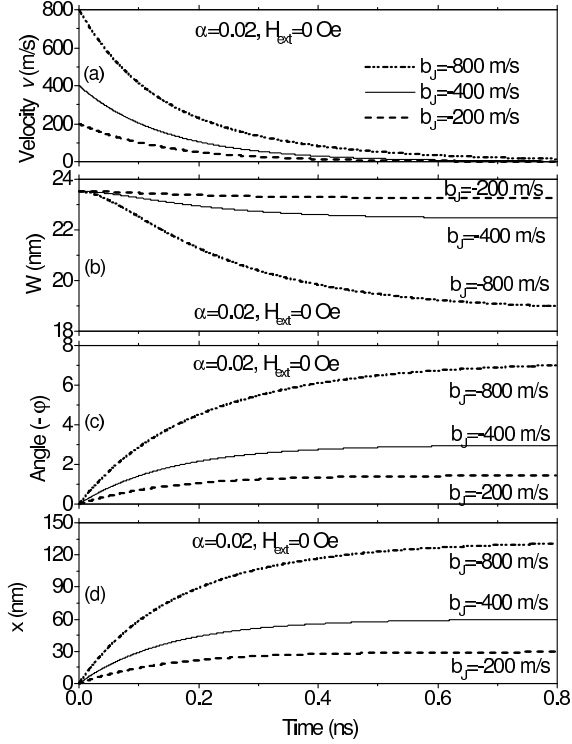}
\caption{The velocity ${\it v}$, the wall width ${\it W}$, the
wall angle $\varphi$ and the displacement $x$ as a function of
time ${\it t}$ for different spin torques and for $H_{ext}=0 Oe$.
The parameters are $4\pi M_s =1.8\times10^{4} {\it Oe}$, $H_K =500
{\it Oe}$, $M_s=14.46\times10^{5} A/m$, $A=2.0\times10^{-11} J/m$
and $\alpha=0.02$.}
\end{figure}

\subsection{Domain wall distortion}

In our analytical theory, domain wall distortion is characterized
by two parameters, $\varphi (t)$ and $c^{-1} (t)$. The former
describes the out-of-plane component of the magnetization and the
latter is the time-dependent domain wall width. When ${\it
t}\rightarrow \infty$, the distortion is maximum. By placing
$d\varphi/dt=0$, we find $\varphi (\infty )$ from Eq.~(14) in the
absence of the external field,
\begin{eqnarray}
\sin2\varphi ( \infty )=\frac{b_{J}c(\infty)}{2\gamma\pi
M_s}=\frac{b_{J}\sqrt{\frac{H_K M_s}{2A}+\frac{4\pi
M_s^2}{2A}\sin^{2}\varphi (\infty ) }}{2\gamma\pi M_s}
\end{eqnarray}
where the last identity is from Eq.~(12). When $b_{J}$ ($\ll\gamma
W_{0}\sqrt{4\pi M_sH_K}$) is small,
$\sin\varphi\approx \varphi$ and $\sin^{2}\varphi\approx 0$,
Eq.~(17) is simplified as,
\begin{equation}
\varphi (\infty ) \approx\frac{b_{J}}{4\pi\gamma M_{s}W_{0}}.
\end{equation}
By inserting this into Eq.~(12) and again by assuming $b_J$ is
small, we find the maximum reduction of the domain wall width is
\begin{equation}
\frac{W (\infty )}{W_0} = 1 -\frac{1}{2W_{0}^{2}}\frac{b_{J}^{2}}{4\pi
M_s H_K \gamma^{2}}
\end{equation}
In Fig.~2 (b) and (c), we show the domain wall distortion during the
application of the current.

\subsection{Domain wall displacement}

As the domain wall motion eventually stops even for the perfect wire
in the absence of the external field, it is interesting to see how far
the domain wall moves, i.e., what is the maximum distance a domain wall
can travel before it stops? We estimate this displacement by integrating
the velocity and utilizing Eqs.~(12) and (13)
\begin{equation}
x_{max} =\int_{0}^{\infty} {\it v} dt
=-\int_{0}^{\varphi (\infty ) }\frac{d\varphi}{\alpha c}
\approx -\frac{b_{J}}{4\pi\gamma\alpha M_{s}}.
\end{equation}
Notice that the displacement is inversely proportional to the damping constant.
In Fig.~2(d), we show the displacement as a function of time.

\subsection{Inductance of domain wall}

We have seen that domain wall motion driven by the current is so
much different from that driven by an external field. In the
latter case, the initial velocity upon the application of the
external field is small and the velocity is gradually increasing
until a saturation velocity is reached; after that, the domain
wall moves with a constant velocity. We can understand this
field-driving wall motion rather straightforwardly: the wall
motion along the direction of the field is to reduce the Zeeman
energy; the rate of the Zeeman energy reduction equals to the
energy damping in a constant moving domain wall. For the
current-driving domain wall motion, the wall motion does not
decrease the energy because the spin torque has no effect on the
{\em uniformly} magnetized domains. Instead, the application of
the spin torque introduces energy to the wall so that distortion
of the wall occurs. The consequence of the wall distortion is to
generate an energy damping mechanism so that the pumping of the
energy by the spin torque can be compensated by the damping. Once
the dynamic equilibrium between pumping and damping is reached,
the domain wall motion stops and the distortion is the maximum. To
quantitatively see this interesting feature, we consider the wall
energy
\begin{equation}
E=\int_{-\infty}^{\infty}\left\{2 \pi M_{z}^{2} -
\frac{H_K}{2M_s}M_{x}^{2} +
\frac{A}{M_{s}^{2}} \left| \frac{\partial {\bf M}}{\partial
x} \right|^{2}\right\}dx
\end{equation}
where the first term is magnetostatic energy, the second is the anisotropy
energy and the final term is the exchange energy. The
rate of the energy change can be obtained via $dE = -{\bf H}_{eff} \cdot
d{\bf M}$ and use the LLG equation,
\begin{eqnarray}
\frac{d E}{d t}
=\int_{-\infty}^{\infty}\left\{-\frac{\gamma\alpha}{1+\alpha^{2}}
\frac{1}{M_s}|{\bf M}\times{\bf
H}_{eff}|^{2}\right.\nonumber \\
\left. -\frac{b_{J}}{1+\alpha^{2}}{\bf H}_{eff}\cdot\frac{\partial
{\bf M}}{\partial x}
-\frac{b_{J}}{1+\alpha^{2}}\frac{\alpha}{M_s}\frac{\partial {\bf
M}}{\partial x} \cdot \left({\bf H}_{eff}\times{\bf
M}\right)\right\}dx\nonumber \\
\end{eqnarray}
Note that the first term is a pure damping, the second and third
terms are the result of the spin torque. By inserting the
effective field Eq.~(4), we can express above two equations in
terms of the wall distortion parameters $\varphi (t)$ and $c(t)$.
The expression is algebraically tedious and lengthy; we do not
write them down here. Instead, we would simply consider limiting
cases where the spin torque is small. A straightforward
calculation leads to simple but insightful expressions for the
rate of the energy change and the wall energy,
\begin{equation}
\frac{d E}{d t} =\frac{16\mu_{0}\pi W_{0} M^{2}_{s}}{3}
\sin\varphi\cos\varphi\frac{d\varphi}{dt}
\end{equation}
and
\begin{equation}
\Delta E \equiv E(b_J) - E(0) = \frac{1}{2} L_{w}b_{J}^{2}
\end{equation}
where we have defined the inductance of the domain wall
$L_{w}=\pi/(3\mu_{0}\gamma^{2} W_{0})$. These expressions
illustrate the energy process during the domain wall motion. The
total spin current induced wall energy is proportional to the
square of the current density ---- an analogy to the inductance of
a circuit.

The stored domain wall energy in the form of domain wall
distortion can be released once the spin current is turned off
---- an analogy to the electric discharge of inductance. In fact, we
can easily show that the domain wall will move back to its
original location. To see this, we set $b_J =0$ in Eq.~(15) and
the velocity for $H_{ext}=0$ is
\begin{equation}
{\it v}_{rec}({\it t})=\frac{\gamma4\pi
M_s\sin\varphi\cos\varphi}{(1+\alpha^2)c({\it t})}
\end{equation}
Noted that the initial condition is now $\varphi=\varphi (\infty
)$ and the initial velocity is $+b_{J}/(1+\alpha^{2})$. By
dropping the first two terms in the right hand side of Eq.~(14),
one can integrate $\varphi (t)$ out from Eq.~(14) and it is easy
to check that
\begin{equation}
x_{rec} \equiv \int_{0}^{\infty} {\it v}_{rec} dt
=\int^{0}_{\varphi (\infty)}\frac{d\varphi}{\alpha c} =-x_{max}
\end{equation}
The domain wall moves to its original position. Again,
this phenomenon is analogous to the discharge of an inductance in
a usual electronic circuit. We will show in the next Section how
this feature differs from the domain wall motion when an external
field is turned off.

\subsection{Beyond Walker's limit}

Up till now, our analysis is based on Walker's trial function. As
in the case of field-driving domain wall motion, it is essential
to assume that the deviation from the original N\'{e}el wall is
small. This places the validity of our analytical result to a
small current density. In this subsection, we show that when the
current density is larger than a critical current, Walker's trial
function fails to describe the domain wall motion. In fact, the
domain wall will not stop at a sufficient large current and the
distortion of the domain wall can not be simply described by two
parameters ($\varphi (t)$ and $c(t)$) anymore. Fortunately, for
the experimental interesting, the applied current density is
usually within the applicability of the Walker's theory.
\begin{figure}
\centering
\includegraphics[width=7.5cm]{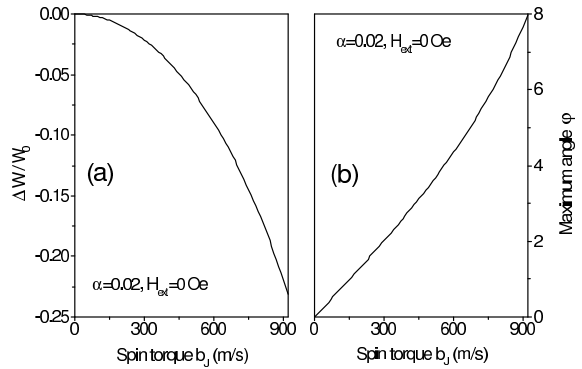}
\caption{Maximum wall width narrowing $\Delta W/W_{0}$ and the
maximum angle $\varphi$ as a function of the spin torque.}
\end{figure}

To find the critical current density, $b_{c}$, we consider a time independent
solution by setting $\frac{d\varphi}{d{\it t}}=0$ in Eq.~(14). By using
Eq.~(12), we find
\begin{equation}
b_{J}=\frac{4\pi\gamma M_s\sin\varphi\cos\varphi}{\sqrt{\frac{H_K
M_s}{2A}+\frac{4\pi M_s^2}{2A}\sin^{2}\varphi}}
\end{equation}
Maximizing $b_{J}$ with respect to the angle $\varphi$, we have
\begin{eqnarray}
\sin^{2}\varphi_{c}=\frac{H_K}{2H_K+4\pi M_s}\\
\sin\varphi_{c}\cos\varphi_{c}=\frac{\sqrt{H_K(H_K+4\pi
M_s)}}{2H_K+4\pi M_s}
\end{eqnarray}
Inserting Eqs.~(28) and (29) back into Eq.~(27), we find the maximum
spin torque for a stationary solution is
\begin{equation}
b_{c}=\frac{4\pi\gamma M_s\sqrt{H_K+4\pi
M_s}}{\sqrt{\frac{M_s}{2A}(2H_K+4\pi M_s)^{2}+\frac{4\pi
M_s^2}{2A}(2H_K+4\pi M_s)}}
\end{equation}
when $4\pi M_s \gg H_{K}$,
\begin{equation}
b_{c}\approx4\pi\gamma M_s\xi
\end{equation}
where $\xi=\sqrt{A/(4\pi M_{s}^{2})}$ is the exchange length.
Thus when $b_{J}
> b_{c}$, the stationary solution of Walker fails. On the other
hand, if the spin torque is smaller than $b_c$, the maximum
distortion shown in Eq.~(28) is small and thus the Walker's trial
function is valid. For a Co nanowire, $4\pi M_s =1.8\times10^{4}
{\it Oe}$, $H_K =500 {\it Oe}$,
$\gamma=1.9\times10^{7}(Oe)^{-1}s^{-1}$, $M_s=14.46\times10^{5}
A/m$ and $A=2.0\times10^{-11} J/m$, the critical current is about
$b_{c}=922 m/s$ or $j_{e}^{c}=2.2\times10^{10}A/cm^{2}$. This
current density is at least one to two orders of magnitude larger than typical
experimental values and thus the trial function and its solutions
are applicable to most of the experiments. In Fig.~3, we show the
maximum deformation of the domain wall; even for the current
density as high as $2\times10^{10} A/cm^2$, the maximum angle
$\varphi$ is about 8 degrees and the wall only shrinks by 22\%.
Thus we confirm that the trial functions are the good
approximation for domain wall dynamics at least in one-dimension
model presented in this section.

\section{Numerical results}

Up till now, our description of domain wall dynamics is based on
the assumption that the out-of-plane angle $\varphi$ is
time-dependent but spatially independent. In a typical nanowire,
the magnetization varies along the direction of the wire width,
i.e., one may need to go beyond 1-d model. Furthermore, a
realistic wire is expected to contain various pinning centers and
it would be interesting to see the effect of pinning on domain
wall motion. In these cases, a micromagnetic calculation is
required to quantitatively investigate the domain wall dynamics.
Here we take an example of a Co nanowire. To make our discussions
concrete, we specify the wire dimension: the thickness is $d=5$
nm, the width is $D=200$ nm and the length is 1.2 $\mu$m. Clearly,
this choice preserves one-dimensional feature of the wire because
$D$ is larger than the domain wall width of Co. The reason for
choosing this example is to make this numerical calculation
directly comparable with the analytical results obtained in the
last section. The magnetic wire is divided into rectangular cells
whose dimension is $2\times 2\times 5 nm^{3}$. For our film with
the large width $D$, the magneto-dipolar interactions between the
cells can be discarded. The wall dynamics was calculated using a
fourth order predictor-corrector method with an 0.8 ps time step.
The stationary N\'{e}el wall is centered at $x=0$ at $t=0$, a spin
current is turned on along $x$ direction, and the external field
is applied along $x$ direction at $t>0$.

\begin{figure}
\centering
\includegraphics[width=7.5cm]{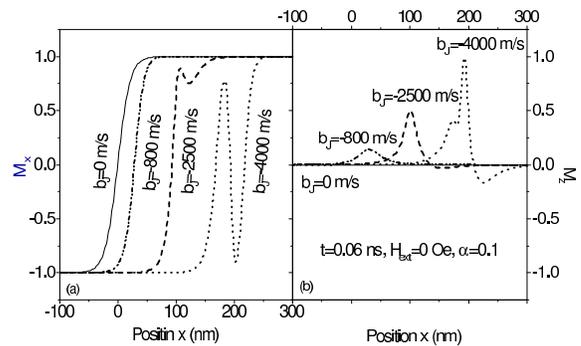}
\caption{The snap shots of the moving magnetic wall at $t=0.06 ns$
in a Co nanowire. The damping constant $\alpha$=0.1 and the
external field $H_{ext}=0 Oe$.}
\end{figure}
In presence of the spin torque, the magnetization of domain wall
is no more confined in the plane of the layer and the original
N\'{e}el wall is distorted during the motion of the domain wall.
In Fig.~4, we show snap shots of magnetization patterns at
$t=0.06$ nanoseconds after a current is turned at $t=0$. For a
small current density, the shape of the N\'{e}el wall is
essentially unchanged except with a small out of plane
magnetization, in consistence with our analytical model in
previous Section. When the applied current is increased to values
larger than the critical current, the shape of the N\'{e}el wall
is completely destroyed. The magnetization inside the wall has
developed a significant out-of-plane component, see Fig.~4b, and
the wall has split into complicated multiple walls. In our
calculation, the critical spin torque $b_{c}\approx 1240 m/s$,
which approximately agrees with our analytical result from
Eq.~(30). We now should confine our discussion below to the case
where the current density is smaller than the critical current.
\begin{figure}
\centering
\includegraphics[width=7.5cm]{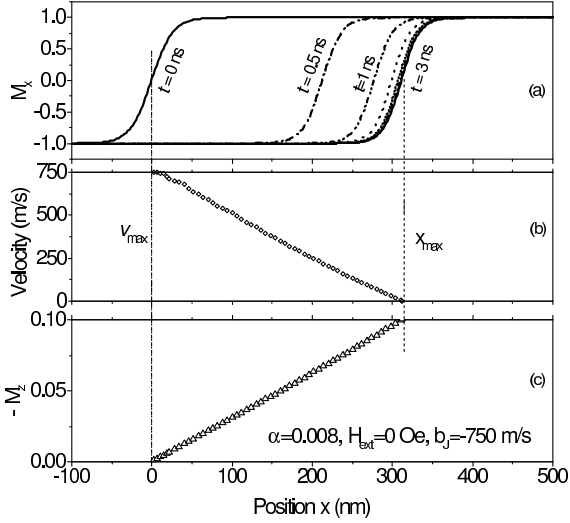}
\caption{(a) The wall positions at several time $t=0$, 0.5ns,
1.0ns, 1.5ns, 2.0ns, 2.5ns  and n=3.0ns; (b) the wall velocity as
a function of the wall position; (c) the maximum out-of-plane
component of the magnetization as a function of the wall position.
The damping constant is $\alpha=0.008$, the spin torque is
$b_{J}=-750 m/s$ and the external field $H_{ext}=0 Oe$.}
\end{figure}
For a perfect wire, the N\'{e}el wall moves with a velocity
$v=-b_J$ immediately after the application of the current. In
Fig.~5, we show wall positions at different times as well as wall
velocity and wall distortion when wall moves to different
positions. The domain wall remains to be a quasi-N\'{e}el wall and
eventually the wall completely stops. At $x=0$, the wall velocity
is at the maximum ($v_{max}=750 m/s$). When the wall stops, the
maximum out-of-plane component of the magnetization
($M_{z}$-component) approaches ($|M_{z}|/M_{s}=0.1$) and the total
displacement is $x_{max}=312 nm$.
\begin{figure}
\centering
\includegraphics[width=7.5cm]{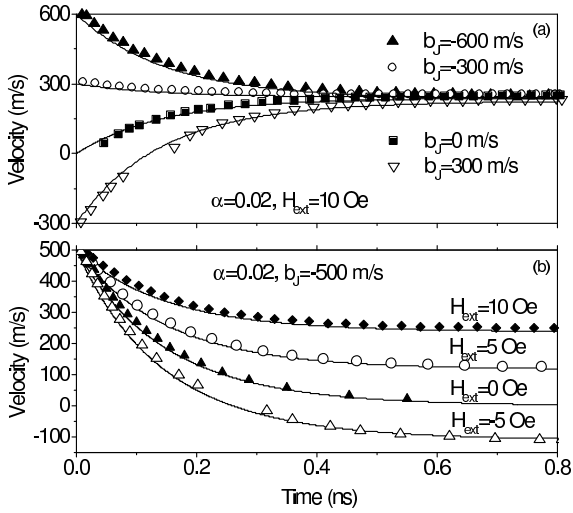}
\caption{Domain wall velocity as a function of time, (a) for
different spin torques at a fixed magnetic field (10 Oe), and (b)
for different magnetic fields at a fixed spin torque ($b_J =-500$
m/s). The solid curves represent the analytical results and
scattered points are the numerical solutions.}
\end{figure}

It is interesting to take a look at the domain wall dynamics with
simultaneous application of the current and the external field. As
we already pointed out that the initial domain wall velocity is
determined by the current but the final terminated velocity is
controlled by the external field, see Fig.~6. We confirm here that the spin
torque alone is unable to move the domain wall over a long distance, see
fig.~7(a). This
\begin{figure}
\centering
\includegraphics[width=7.5cm]{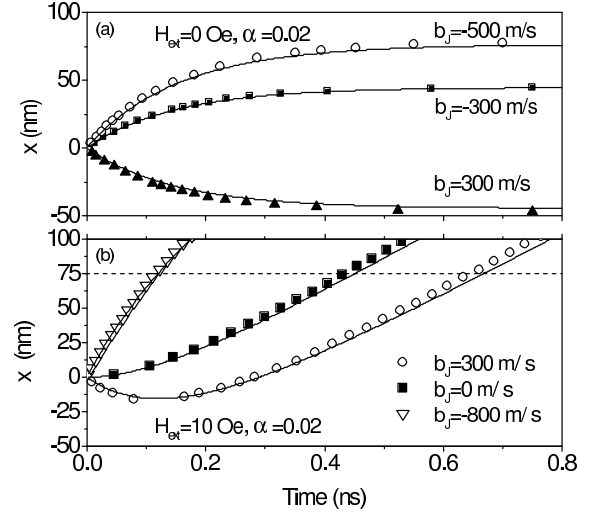}
\caption{Displacement of the domain wall for several spin torques
without (a) and with (b) the external field. The solid curves
represent the solutions of the trial function and scattered points
are the numerical results.}
\end{figure}
feature is just opposite to domain wall motion driven by an
external field: at an initial application of the magnetic field,
the wall velocity is small (in fact one can easily show it is
$\alpha H_{ext} W_0/(1+\alpha^{2})$ from Eq.~(15)), but it becomes
faster and faster until it reaches a saturated speed from Eq.~(16)
at $v_s =\gamma H_{ext}W(\infty)/\alpha$. The origin of this
difference had been discussed in great length in the last section
and we again confirm that these conclusions remain valid for the
realistic wire.
\begin{figure}
\centering
\includegraphics[width=7.5cm]{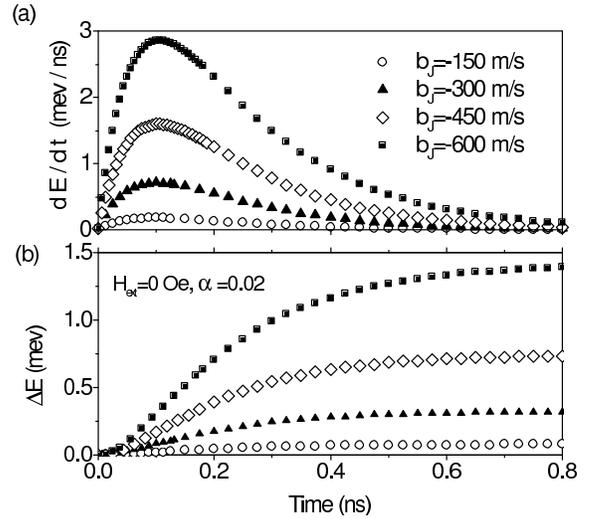}
\caption{Energy power (a) and net energy gain (b) of the domain
wall as a function of time. }
\end{figure}
The fact, that the initial domain wall velocity is determined by
the current but the final terminated velocity is controlled by the
external field, can be used to design the high/low-speed domain
wall propagation by properly choosing the desired spin current
density and the magnetic field. For example, if we are to move the
domain wall by 75 nanometers, the magnetic field (10 $Oe$) alone
will take 0.43 nanoseconds. With help of the spin torque
$b_J=-800m/s$, the same displacement can be made in 0.11
nanoseconds and it takes 0.65 nanoseconds when $b_J=300m/s$. These
features are clearly illustrated in Fig.~7(b).

We show the rate of energy input of the domain wall during the
application of the spin torque in Fig.~8. The rate of energy
pumping from the spin torque increases at initial application of
the spin current, then it develops a maximum rate. After that, the
rate decreases due to increase of the damping. Finally, the energy
pumping is exactly compensated by the damping and the wall motion
stops.

It is interesting to compare our calculations with experiment
\cite{Shinjo2} where a domain wall speed about $3 m/s$ for a
current density $1.2 \times 10^8 A/cm^2$ was observed. As we have
concluded that the spin torque alone is unable to move the wall
over long distance ($1 \mu$m), the observed domain wall velocity
over large distance must be from two possible scenarios: 1) An
additional magnetic field generated by the current (Oersted
fields) was present in the experiment. One just needs a very small
magnetic field to generate this low velocity domain wall motion.
In fact, one to two Oersted was sufficient and it was hard to rule
out such small field in experiment. 2) There is a possibility of
other spin-current induced torque which has different form as
Eq.~(2). Indeed, if one relaxes the adiabatic process that assumes
the direction of the spin polarization of the current
adiabatically follows the magnetization vector in the domain wall,
one would generate other spin torque. The detail calculation of
the spin torque from non-adiabatic process is beyond scope of the
present work; we will defer to a future publication. Other
experiments \cite{Fert2,Parkin}, reported that a magnetic field is
always required to move the wall over a long distance. These
experiments agree with our general conclusions.
\begin{figure}
\centering
\includegraphics[width=8cm]{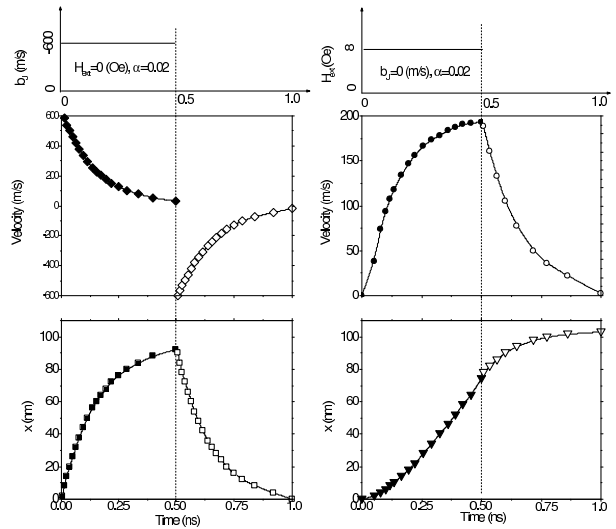}
\caption{Velocity ${\it v}$ and displacement
$x$ for a pulsed spin current (left panels) and for a pulsed
magnetic field (right panels).  The damping constant is $\alpha=0.02$.}
\end{figure}

We end the section by looking at two more interesting examples.
The first one is that the spin current is applied in the form of a
pulse. If we applied a rectangular pulsed spin current along $x$
direction, what dynamic behaviors do the domain walls have? In
Fig.~9, we show the evolution of velocity ${\it v} (t)$ and
displacement $x(t)$ with the pulsed spin toque and we compared
them with the pulsed magnetic field. Suppose the spin torque with
the magnitude $b_{J}=-600 m/s$ is turned on at t=0 and turned off
at $\tau=0.5 ns$. In the absence of the external field, the domain
wall moves back to its original location after the current is
turned off. This is again consistent with our analytical result.
However, for the field-driving domain wall motion, the domain wall
continues to move in the forward direction after the field is
turned off.

Another example is to model the effect of pinning centers in the
domain wall motion. In realistic nanowires, domain walls are not
completely free to move. There are various pinning sources such as
defects and roughness. We introduced a pinning field,
$H_{pin}=V_{0}x H(\zeta-|x|)/\zeta$, where $H(y)$ is the Heaviside
step function, $\zeta>0$ is the distance from the domain wall
center, $x$ is the position.
\begin{figure}
\centering
\includegraphics[width=7.5cm]{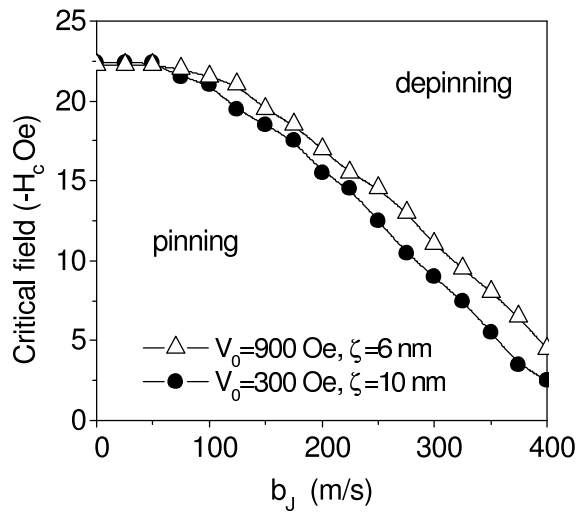}
\caption{The phase boundary of pinning and depinning domain wall
for two types of pinning centers, see text for the definition of
the pinning parameters $V_0$ and $\zeta$. The external field is
applied along the $-x$ direction and the spin current is along
$+x$ direction at $t=0$.  The damping constant is $\alpha=0.02$.}
\end{figure}
In Fig.~10, we present the result for the critical magnetic field
required to move the wall in the presence of the defect. The role
of the current is to first move the domain wall out of the pinning
center so that the critical magnetic field required to move the
wall is smaller.

\section{Conclusions}

We have demonstrated in this paper that the domain wall motion
driven by a spin current has many unique features that do not
exist in the conventional domain wall motion driven by a magnetic
field. By solving the Landau-Lifshitz-Gilbert equation along with
the current induced spin torque, we summarize below our main
findings that are supported by our analytical and numerical study:
1) the spin torque alone can move the domain wall, but the total
displacement is limited, thus the spin torque is not capable to
move domain walls over a large distance; 2) since the speed of the
wall is large at an initial application of the spin torque, there
is an advantage of the spin torque over the magnetic field for
wall movement at short distance; 3) the domain wall is capable to
store current-induced energy and thus the energy process of the
domain wall is very similar to the charge/discharge of a
inductance in an electric circuit; 4) the spin torque can help to
reduce the critical field needed to move domain wall if domain
wall pinning centers are present; 5) it may be important to extend
a spin torque other than adiabatic spin torque used here in order
to study the contribution from the non-adiabatic spin torque whose
form has not yet been investigated.

In conclusion, a spin torque with the form of
Eq.~(2) is proposed to study the dynamics of the domain wall. Due
to different roles played by the spin torque and by the external
field, the current driven domain wall motion represent new types of
torques. This new spin torque creates opportunity for a fast domain wall
motion by using the combined field and spin torque. The research
was supported by NSF (ECS-0223568 and DMR-0314456).

{\em Note added in proof.} a paper was published,{\cite Tatara }
in which the critical current of domain wall motion was briefly
discussed.


\begin{thebibliography}{1}

\bibitem{Buhrman}
J. A. Katine, F. J. Albert, R. A. Buhrman, E. B. Myers, and D. C.
Ralph, Phys. Rev. Lett. {\bf 84}, 3149 (2000).

\bibitem{Fert1}
J. Grollier, V. Cros, A. Hamzic, J. M. George, H. Jaffres, A.
Fert, G. Faini, J. Ben Youssef, and H. Legall, Appl. Phys. Lett.
{\bf 78}, 3663 (2001).

\bibitem{Kent}
B. \"{O}zyilmaz, A. D. Kent, D. Monsma, J. Z. Sun, M. J. Rooks,
and R. H. Koch, Phys. Rev. Lett. {\bf 91}, 067203 (2003).

\bibitem{Bass1}
S. Urazhdin, N. O. Birge, W. P. Pratt Jr., and J. Bass, Phys. Rev.
Lett. {\bf 91}, 146803 (2003).

\bibitem{Tsoi}
M. Tsoi, V. Tsoi, J. Bass, A. G. M. Jansen, and P. Wyder, Phys.
Rev. Lett. {\bf 89}, 246803 (2002).

\bibitem{Myers}
S. I. Kiselev, J. C. Sankey, I. N. Krivorotov, N. C. Emley, R. J.
Schoelkopf, R. A. Buhrman, D. C. Ralph, Nature  {\bf 425}, 380
(2003).

\bibitem{Sorb}
For a review see. e.g., R. S. Sorbello, {\it Solid State Physics},
H. Ehrenreich and F. Spaepen eds. (Academic Press, New York,
1998), Vol {\bf 51}, p. 159, and references therein.

\bibitem{Slon}
J. Slonczewski, J. Magn. Magn. Mater. {\bf 159}, L1 (1996);{\bf
195}, L261(1999).

\bibitem{Waintal}
X. Waintal, E. B. Myers, P. W. Brouwer, and D. C. Ralph, Phys.
Rev. B {\bf 62}, 12317(2000).

\bibitem{Stiles2}
M. D. Stiles and A. Zangwill, Phys. Rev. B {\bf 66}, 014407(2002).

\bibitem{Brataas}
A. Brataas, Y. V. Nazarov, and G. E. W. Bauer, Phys. Rev. Lett.
{\bf 84}, 2481(2000).

\bibitem{Levy}
S. Zhang, P. M. Levy and A. Fert, Phys. Rev. Lett. {\bf 88},
236601 (2002).

\bibitem{Sun2}
J.Sun, Phys. Rev. B {\bf 62}, 570(2000).

\bibitem{Bazaliy2}
Ya. B. Bazaliy, B. A. Jones, and S.-C. Zhang, J. Appl. Phys. {\bf
89}, 6793 (2001).

\bibitem{Miltat3}
J. Miltat, G. Albuquerque, A. Thiaville, and C. Vouille, J. Appl.
Phys. {\bf 89}, 6982(2001).

\bibitem{Li2}
Z. Li and S. Zhang, Phys. Rev. B {\bf 68}, 024404 (2003); Phys.
Rev. B {\bf 69}, 134416 (2004).

\bibitem{Berger}
L. Berger, J. Appl. Phys. {\bf 49}, 2156 (1978); {\bf 55}, 1954
(1984).

\bibitem{Bazaliy}
Ya. B. Bazaliy, B. A. Jones, and S.-C. Zhang, Phys. Rev. B {\bf
57}, R3213 (1998).

\bibitem{Zhang}
Z. Li and S. Zhang, Phys. Rev. Lett. {\bf 92}, 207203 (2004).

\bibitem{Walker}
N. L. Schryer and L. R. Walker, J. Appl. Phys. {\bf 45}, 5406
(1974).

\bibitem{Shinjo2}
A. Yamaguchi, T. Ono, S. Nasu, K. Miyake, K. Mibu, and T. Shinjo
Phys. Rev. Lett. {\bf 92}, 077205 (2004) .

\bibitem{Fert2}
J. Grollier, P. Boulenc, V. Cros, A. Hamzi\'{c}, A. Vaur\'{e}s, A.
Fert and G. Faini, Appl. Phys. Lett. {\bf 83}, 509 (2003).

\bibitem{Parkin}
M. Tsoi, R. E. Fontana, S. S. P. Parkin, Appl. Phys. Lett. {\bf
83}, 2261 (2003).

\bibitem{Tarata}
G. Tatara and H. Knohno, Phys. Rev. Lett. {\bf 92}, 086601 (2004).

\end{thebibliography}
\end{document}